\documentstyle[11pt, aaspp4, epsf, rotate]{article}

\begin{document}

\title{Line emission from gamma-ray burst environments}

\author{Markus B\"ottcher\footnote{Chandra Fellow}\altaffilmark{,2}}

\altaffiltext{2}{Department of Space Physics and Astronomy, 
Rice University, MS 108, 6100 S. Main St., Houston, TX 77005-1892}

\centerline{\it Submitted to the Astrophysical Journal}

\begin{abstract} 
The time and angle dependent line and continuum emission 
from a dense torus around a cosmological gamma-ray burst 
source is simulated, taking into account photoionization, 
collisional ionization, recombination, and electron 
heating and cooling due to various processes. The importance 
of the hydrodynamical interaction between the torus and 
the expanding blast wave is stressed. Due to the rapid 
deceleration of the blast wave as it interacts with the dense 
torus, the material in the torus will be illuminated by a 
drastically different photon spectrum than observable through 
a low-column-density line of sight, and will be heated by
the hydrodynamical interaction between the blast wave and
the torus. A model calculation to reproduce the Fe~K$\alpha$ 
line emission observed in the X-ray afterglow of GRB~970508 
is presented. The results indicate that $\sim 10^{-4} \, 
M_{\odot}$ of iron must be concentrated in a region of 
$R \lesssim 10^{-3}$~pc. The illumination of the torus
material due to the hydrodynamical interaction of the blast
wave with the torus is the dominant heating and ionization
mechanism leading to the formation of the iron line. These
results suggest that misaligned GRBs may be detectable as
X-ray flashes with strong emission line features.
\end{abstract}


\section{Introduction}

The recent marginal detections of redshifted iron
K$\alpha$ emission lines in the X-ray afterglows of
GRB~970508 (\cite{piro99}) and GRB~970828 
(\cite{yoshida99}) have stimulated vital interest 
in the processes of photoionization and fluorescence 
line emission in gamma-ray burst (GRB) environments
as a potentially powerful tool for redshift measurements
and mapping of the density structure and composition 
of the vicinity of cosmological GRBs. X-ray absorption 
features and fluorescence line emission from the 
environments of cosmological GRBs have been investigated 
theoretically by M\'esz\'aros \& Rees (\markcite{mr98b}1998b), 
B\"ottcher et al. (\markcite{boettcher99a}1999a),
B\"ottcher, Dermer \& Liang (\markcite{boettcher99b}1999b),
and Ghisellini et al. (\markcite{ghi99}1999). These 
studies were motivated by suggestions that GRBs 
are caused by the death of a very massive star 
(\cite{woosley93}; \cite{pac98}; \cite{sv98}) and are 
therefore likely to be embedded in the dense gaseous 
environment of a star-forming region. The {\it HST} image of the 
optical counterpart of the extremely bright GRB~990123 
possibly provided the first direct observation of an 
association of a GRB with a star-forming region of the 
proposed host galaxy (\cite{bloom99a}). (Note, however,
that Holland \& Hjorth [\markcite{hh99}1999] have found 
that the optical counterpart is located at a distance of
$\gtrsim 1.3$~kpc from the nearest star-forming region
resolvable in the {\it HST} image of the host galaxy after 
subtraction of the optical transient, which is larger 
than the expected separation of a very massive star from 
its birth place.) Bloom et al. (\markcite{bloom99a}1999a) have 
also demonstrated that those GRBs with optical afterglows 
and identification with a host galaxy are significantly offset 
from the centers of their host galaxies, but generally inside 
the galaxy's effective radius. The possible association of 
GRB~980425 with the Type Ic supernova SN~1998bw (\cite{galama98}) 
and the marginal evidence of a supernova underlying the optical 
light curve of GRB~980329 (\cite{bloom99b}) provide further
support for GRB source models related to the death of massive 
stars, at least for those $\sim 50 \%$ of well-localized GRBs 
for which optical afterglows are observable and the correlation 
with host galaxies can be established. 

For this reason, Ghisellini et al. (\markcite{ghi99}1999)
and B\"ottcher et al. (\markcite{1999a}1999a) have investigated
the influence of a dense, quasi-isotropic GRB environment, as
appropriate for a star-forming region, on the observable 
radiation, in terms of X-ray absorption features and fluorescence
line emission. However, the results of B\"ottcher et al. 
(\markcite{boettcher99}1999a, \markcite{boettcher99b}b) 
indicate that while a temporally varying Fe~K absorption 
edge might be a powerful tool to gain information even about 
an isotropic GRB environment, the luminosity and duration of 
the Fe~K$\alpha$ line observed in GRB~970508 (\cite{piro99}), 
if real, is inconsistent with a quasi-isotropic environment. 
This is mainly because the column density of material along 
our line of sight toward the GRB is strongly constrained by 
the lack of detectable X-ray absorption features and by the 
requirement that the Thomson scattering optical depth 
$\tau_{\rm T}$ along the line of sight must be $\tau_{\rm T} 
\ll 1$ because otherwise the short-term variability of the 
burst radiation would be smeared out over the photon escape 
time scale from the environment.

A plausible way to solve this problem is the assumption
of a strongly anisotropic GRB environment from which a
GRB is only observed if we happen to look at the system
through a line of sight with low absorption and Thomson
depth. Such a geometry may be provided in the form of
a dense torus surrounding the GRB source which could
be produced by anisotropic ejecta of the burst progenitor,
such as a dense stellar wind ejected by the progenitor of 
a hypernova (\cite{pac98}) or the remnant of a supernova 
preceding a supranova (\cite{sv98}). That equatorial density
enhancements around massive stars and supernova remnants may be 
a common phenomenon, is also supported by direct observations, e. g. 
of the bipolar structure of $\eta$ Carinae (\cite{dh97}) and the 
torus-shaped geometry visible in the {\it Chandra} image of this 
star\footnote{see http://chandra.harvard.edu/photo/0099/index.html}, 
the double ring structure of the remnant of SN~1987A, and the 
recent {\it Chandra} image of the central regions of the Crab 
nebula\footnote{see http://chandra.harvard.edu/photo/0052/index.html}. 

An anisotropic geometry has recently been considered by 
Lazzati et al. (\markcite{lcg99}1999), who qualitatively 
discussed several relevant physical processes 
potentially leading to an observable Fe~K$\alpha$ line
in the X-ray afterglow of a GRB. The processes considered
by these authors include (1) fluorescence line emission
following multiple photoionization events, with the
efficiency of line emission determined by the time scale
for re-filling of the inner atomic shells of Fe by 
recombination in a very dense environment, (2) recombination
and fluorescence line emission following electron-collisional 
ionization in a dense, hot plasma (called ``thermal emission'' 
in \cite{lcg99}), (3) fluorescence line emission 
in the course of reflection of the GRB radiation 
off a dense, highly opaque medium, in which most 
of the iron remains in a low ionization state. They 
find that the multiple-ionization -- recombination scenario
has problems due to the high electron temperature and
implied long recombination times expected under the
conditions where this process might be dominant, while
their ``thermal emission'' and ``reflection'' scenarios
appear more promising. Both processes require a very
dense, highly opaque torus illuminated and heated by
the GRB radiation. Thus, under these conditions, both
processes might be important to a certain extent.

Recently, Weth et al. (\markcite{weth00}2000) have studied
the radiative response of an anisotropic GRB environment to
the illumination by the prompt and afterglow radiation of
the GRB, assuming that at any given point in time the radiation
transport can be treated in thermal and ionization equilibrium,
and that the blast wave associated with the GRB does not
interact hydrodynamically with the dense material responsible
for potentially observable absorption and emission features.
In this approximation, they exploit a quasi-isotropic shell
geometry and the geometry of an evacuated funnel, the walls
of which act as a warm (partially ionized) Compton reflector. 
While the shell geometry was found to produce predominantly
absorption features, followed by weaker line emission than 
observed in GRB~970508, the funnel geometry was appropriate
to reproduce the observed Fe~K$\alpha$ line from that burst.

In this paper, I present a detailed, time-dependent numerical 
study of the relevant processes in a dense torus illuminated 
by GRB radiation. The code used for this study originated 
from the time-dependent photoionization and line transfer 
code used in B\"ottcher et al. (\markcite{boettcher99a}1999a, 
\markcite{boettcher99b}b). The physics included in the 
numerical study is discussed in \S 2. In \S 3, general 
considerations and an appropriate model setup to reproduce 
the observed afterglow continuum of GRB~970508 
(\cite{piro98},\markcite{piro99}1999) are presented. General
results of a parameter study, from which the required 
torus parameters necessary to reproduce the (marginally) 
detected Fe~K$\alpha$ line in the X-ray afterglow of GRB~970508 
(\cite{piro99}) can be deduced, are discussed in \S 4. 
I summarize in \S 5. 

\section{Model assumptions and computational scheme}

In a previous paper (\cite{boettcher99a}), we have 
investigated in detail the problem of time-dependent 
photoionization, photoelectric absorption and fluorescence 
line emission in the case of an isotropic, moderately dense 
GRB environment. Under the conditions investigated in that 
paper, recombination and electron-collisional effects were 
negligible and were thus not included in our treatment. In
the situation investigated in this paper, such effects have
to be considered carefully. Furthermore, the numerical problem 
is no longer isotropic. The code used in B\"ottcher et al. 
(\markcite{boettcher99a}1999a) has thus been modified in order
to account for the anisotropy of the GRB environment. 

\begin{figure}
\epsfysize=9cm
\epsffile[0 100 450 450]{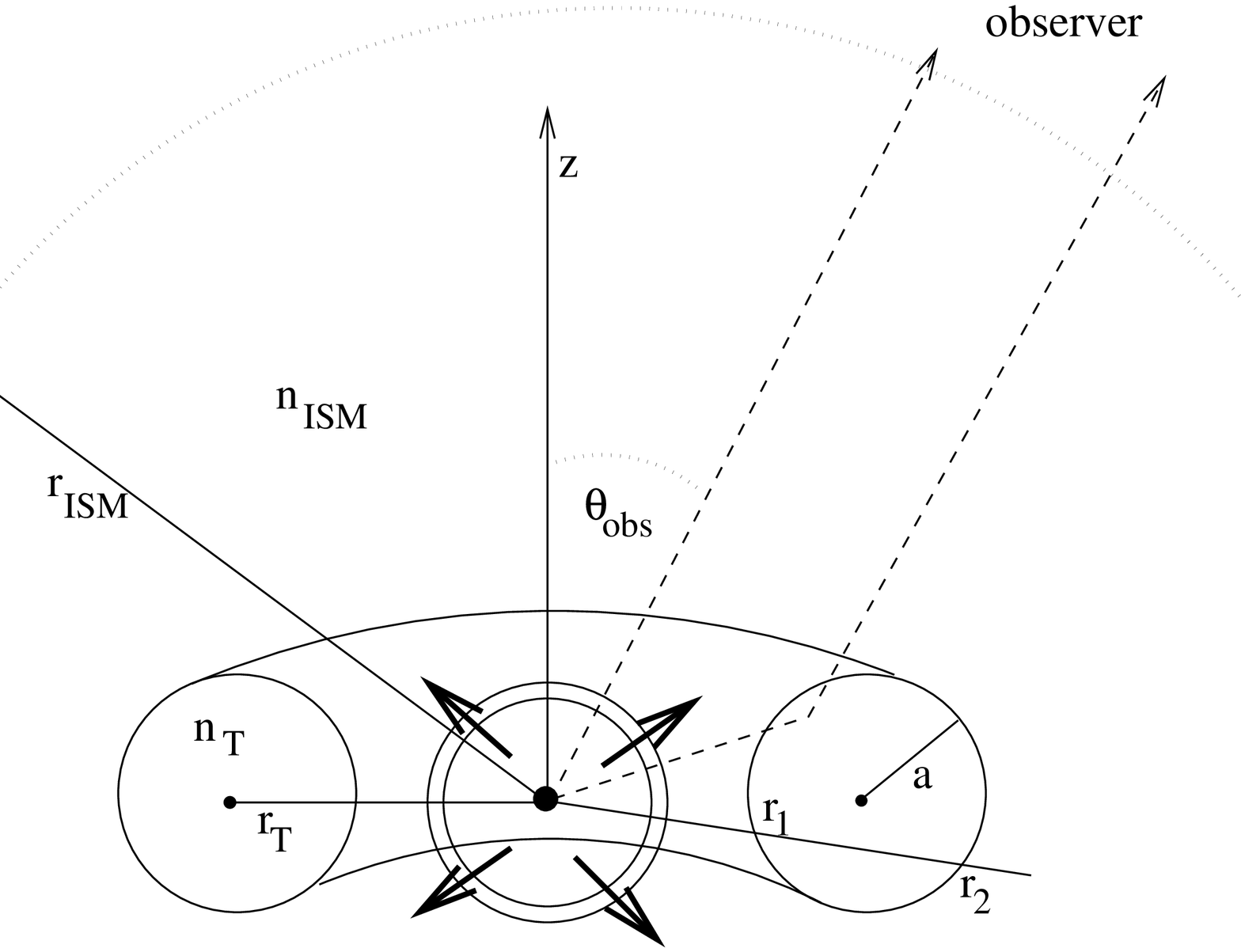}
\caption[]{Illustration of the model geometry: A relativistic blast
wave, initiated at the origin, interacts with the surrounding
medium and emits the observable GRB and afterglow radiation. The
burst source is surrounded by a torus of dense, metal enriched
material at a distance $r_{\rm T}$ from the center of the explosion.
$a$ is the cross-sectional radius of the torus. The entire configuration 
is embedded in a homogeneous ISM with density $n_{\rm ISM}$ and standard
solar-system element abundances.}
\label{geometry}
\end{figure}

The geometry assumed to treat this problem is illustrated in
Fig. \ref{geometry}. The center of the coordinate system is the 
center of the GRB explosion. The GRB source is surrounded by a 
torus of dense material (particle density $n_{\rm T}$), at a 
distance $r_{\rm T}$ from the center of the explosion. The 
radius of the cross-section of the torus is denoted by $a$. 
The burst source and the torus are embedded in a dilute ISM 
of density $n_{\rm ISM} (r)$ which extends out to a radius 
$r_{\rm ISM}$ from the center of the explosion and is assumed
to be distributed isotropically, but generally not homogeneously
around the source. 

The ISM and the torus are illuminated by the time-dependent
radiation field of a GRB. The temporal evolution of the GRB
radiation field is represented by the analytical expressions
of Dermer, Chiang \& B\"ottcher (\markcite{dcb99}1999), 
approximating the detailed predictions of external synchrotron 
shock model for GRBs (\cite{rm92}; \cite{mr93}; \cite{katz94}; 
\cite{tavani96}). The numerical scheme used to treat the 
time-dependent radiation transport problem is basically the 
same as described in B\"ottcher et al. 
(\markcite{boettcher99a}1999a), except that now the 
environment is anisotropic and more processes are included. 
The environment is split up in several (up to 22) angular 
zones, within which radial transfer of the GRB radiation is 
considered. To account for the anisotropy of the circumburster
material (CBM), the expressions of Dermer et al. 
(\markcite{dcb99}1999) for the blast wave evolution need to be 
modified because the blast wave will be decelerated much more 
efficiently in the dense torus than in the dilute ISM. Each 
angular element of the CBM will be treated separately as if 
it were part of an isotropic, but inhomogeneous environment, 
characterized by a spherical shell of dense material of 
uniform density $n_{\rm T}$ extending from an inner radius 
$r_1$ to an outer radius $r_2$ (see Fig. \ref{geometry}). 
If the effect of blast wave deceleration in the dilute 
ISM prior to hitting the dense torus is negligible, the 
effective deceleration radius $r_{\rm d}$ will now be located 
at

\begin{equation}
r_{\rm d} = \left( {3 \, E_0 \over 4 \pi \, n_{\rm T} \, m_{\rm p} c^2 
\, \Gamma_0^2} + r_1^3 \right)^{1/3},
\label{xd}
\end{equation}
where $E_0$ is the total energy of the primary ejecta in the
spherical blast wave, $\Gamma_0$ is the bulk Lorentz factor of 
the material behind the shock front, and $m_{\rm p}$ is the proton
mass. The deceleration of the blast wave may then be parametrized 
as a decrease of the bulk Lorentz factor $\Gamma$ with radius 
$r_{\rm b}$ of the blast wave from the center of the explosion:

\begin{equation}
\Gamma(r_{\rm b}) = \Gamma_0 \cdot \cases{ 1 & if $r_{\rm b} < r_{\rm d}$, \cr
                    \xi^{-g} & if $r_{\rm b} \ge r_{\rm d}$, \cr}
\label{Gamma}
\end{equation}
where

\begin{equation}
\xi \equiv \left( {r_{\rm b}^3 - r_1^3 \over r_{\rm d}^3 - r_1^3} 
\right)^{1/3}.
\label{xi}
\end{equation}
The photon energy of the $\nu F_{\nu}$ spectrum will shift
according to

\begin{equation}
\epsilon_{\rm p} (r_{\rm b}) = \epsilon_0 \, \left( {n_{\rm T} 
\over n_{\rm ISM}} \right)^{1/2} \cdot 
\cases{ {\rm const.} & if $r_1 < r_{\rm b} < r_{\rm d}$, \cr
          \xi^{-4 g} & if $r_{\rm b} \ge r_{\rm d}$, \cr}
\label{ep}
\end{equation}
where $\epsilon_0$ is the photon energy of the $\nu F_{\nu}$ spectrum
at the peak of the burst in a direction where the blast wave only
interacts with the dilute ISM. The $\nu L_{\nu}$ peak luminosity is

\begin{equation}
P_{\rm p} (r_{\rm b}) = 2 \pi \, c \, {2 g - 3 \over g} \, {m_{\rm p} c^2 
\, \Gamma_0^4 \, n_{\rm T} \over (\upsilon^{-1} + \delta^{-1})} \>
r_{\rm b}^2 \cdot \cases{ {\rm const.} & if $r_1 < r_{\rm b} < r_{\rm d}$, \cr
                      \xi^{-4 g} & if $r_{\rm b} \ge r_{\rm d}$, \cr}
\label{Pp}
\end{equation}
where $\upsilon$ and $\delta$ are the spectral indices,
$\nu F_{\nu} \propto \epsilon^{\upsilon}$ below and 
$\nu F_{\nu} \propto \epsilon^{-\delta}$ above the peak
photon energy, respectively.

Inspection of eqs. \ref{xd} -- \ref{Pp} shows that the interaction
of the blast wave with the torus leads to an extremely short, 
extremely luminous flash of very high-energy radiation. The 
duration of this flash, as it would be measured by an observer 
located within or behind the torus, is given by

\begin{equation}
\Delta t_{\rm t} = {r_{\rm d} - r_1 \over 2 \, \Gamma_0^2 c} \approx
{E_0 \over 8 \pi \, n_{\rm T} \, m_{\rm p} c^3 \, \Gamma_0^4 \, r_1^2}
\nonumber \\
\approx 1.1 \times 10^{-5} \, {\rm s} \> {E_{54} \over n_{10} \,
\Gamma_{300}^4 \, r_{15}^2},
\label{dt_t}
\end{equation}
where $E_{54} = E_0 / (10^{54} \, {\rm ergs})$, $n_{10} = n_{\rm T} 
/ (10^{10} \, {\rm cm}^{-3})$, $\Gamma_{300} = \Gamma / 300$, and
$r_{15} = r_1 / (10^{15} \, {\rm cm})$.

As the parts of the blast wave interacting with the torus are 
decelerated to subrelativistic velocities almost instantaneously, 
the material of the torus will be energized via shock heating and
via photoionization by the X-ray and gamma-ray flash emitted 
during the phase of blast wave deceleration in the torus. The
heating due to the subrelativistic shock is taken into account 
using basic energy and momentum conservation arguments. As has 
been argued by Vietri et al. (\markcite{vietri99}1999), the 
shock-heated torus material will attain a temperature of
$\sim 3 \cdot 10^{7}$~K, leading to strong Fe~K$\alpha$ line 
emission via electron-impact ionization and subsequent 
recombination.

Line and continuum emission resulting from atomic processes 
in the ISM and the torus is assumed to be emitted isotropically 
at each point. The time delay (due to the light travel time 
difference) of such radiation reaching the observer from 
directions misaligned with respect to the line of sight to 
the GRB source, is properly taken into account. Since the 
column density of ISM and torus material along the line of 
sight as well as the light-travel-time delays depend on the
viewing angle, the output spectra and light curves are sampled 
under different viewing angles $\theta_{\rm obs}$ with respect 
to the symmetry axis of the torus, which defines the $z$ axis.

In addition to the processes of photoelectric absorption,
photoionization and fluorescence line emission following
photoionization events, which had been included already in
B\"ottcher et al. (\markcite{boettcher99a}1999a), now radiative
and dielectronic recombination, electron-collisional ionization,
electron heating and cooling due to bremsstrahlung, Compton 
scattering, electron-impact ionization and Coulomb scattering, 
and continuum emission due to radiative recombination and 
bremsstrahlung emission are taken into account. 

The radiative and dielectronic recombination rates are 
calculated using the results of Nahar \& Pradhan 
(\markcite{np97}1997) for C and N, of Nahar 
(\markcite{nahar99}1999) for O, and of Nahar, Bautista 
\& Pradhan (\markcite{nbp97}1997, \markcite{nbp00}2000), 
Nahar (\markcite{nahar96}1996, \markcite{nahar97}1997),
and Nahar \& Bautista (\markcite{nb99}1999) for Fe 
I -- V. The recombination rates for C-like ions are
taken from Nahar (\markcite{1995}1995), and for Si-like
ions from Nahar (\markcite{naher00}2000). For the 
remaining iron ions, we use the tables of Arnaud \& 
Raymond (\markcite{ar92}1992). For ions in the He, 
Li, Be, and Ne isoelectronic sequences and not covered 
by the above references, we use the recombination rates 
of Romanik (\markcite{romanik88}1988). For the remaining 
ions, we use the coefficients given by Aldrovandi \& 
P\'equignot (\markcite{ap73}1973), and Shull and van 
Steenberg (\markcite{shull82}1982). A few subroutines of 
the XSTAR code (\cite{kallman82}) have been used. In addition 
to the corrections published in the Erratum to Shull and 
van Steenberg (\markcite{shull82}1982), the following 
entries of their table 2 need to be corrected (Shull 
1999, private communication):
$A_{\rm di} ({\rm Si \> IX}) = 4.25 \cdot 10^{-2}$, 
$A_{\rm rad} ({\rm Si \> XI}) = 1.20 \cdot 10^{10}$,
$A_{\rm di} ({\rm Ca \> IX}) = 4.02 \cdot 10^{-2}$,
$A_{\rm rad} ({\rm Ca \> XI}) = 8.51 \cdot 10^{-11}$,
$T_0 ({\rm Fe \> XVI}) = 8.18 \cdot 10^6$,
$T_1 ({\rm Fe \> XXIV}) = 1.17 \cdot 10^7$.
The rates of recombination into excited states are 
calculated using the hydrogenic approximation (Seaton 
\markcite{seaton59}1959). $133$ strong UV and X-ray 
lines due to radiative transitions following 
recombination into excited states been included 
using the line energies and branching ratios given 
by Kato (\markcite{kato76}1976).

Due to the very high Compton equilibrium temperatures 
in the GRB surroundings, collisional ionization will 
largely dominate over collisional excitation so that
excited states will be only weakly populated. Thus,
for ease of computation, only photoionization from
atoms and ions in the ground state is included.

The temperature of free electrons is determined by the excess
energy of ionizing photons, the average energy of Auger electrons 
as given by Kaastra \& Mewe (\markcite{km93}1993), Compton 
heating/cooling, Coulomb interactions with cold protons 
(using the prescription of Dermer \& Liang [\markcite{dl89}1989]) 
and bremsstrahlung and collisional ionization energy losses.

The electron-collisional ionization rates are calculated using
the prescription of Shull \& van Steenberg (\markcite{shull82}1982).
The standard formula for electron-ion bremsstrahlung emission 
(\cite{rl79}) and the expression of Tucker \& Koren 
(\markcite{tk71}1971) for recombination continuum emission 
are used.

The elements included in the simulations are H, He, C, 
O, Ne, Mg, Si, S, Ca, Fe, and Ni since these are the most
abundant elements in astrophysical plasmas, for which the 
relevant atomic data are readily available in the literature. 
The contributions from other elements have been neglected. 
For the dilute surroundings of the GRB source, standard
solar-system abundances, from Zombeck (\markcite{zombeck90}1990),
are assumed. Normalized to 1, the relative abundances are 
$0.935$ (H), $6.33 \times 10^{-2}$ (He),
$3.90 \times 10^{-4}$ (C), $8.12 \times 10^{-5}$ (N),
$6.47 \times 10^{-4}$ (O), $9.14 \times 10^{-5}$ (Ne),
$3.73 \times 10^{-5}$ (Mg), $3.52 \times 10^{-5}$ (Si),
$1.76 \times 10^{-5}$ (S), $3.73 \times 10^{-6}$ (Ar),
$2.20 \times 10^{-6}$ (Ca), $3.16 \times 10^{-5}$ (Fe),
$1.68 \times 10^{-6}$ (Ni). 

The chemical composition of the material in the torus is dominated 
by the ejecta of the GRB progenitor. The association of GRB~980425
with the type Ic SN~1998bw suggests that this material may be similar
in composition to the ejecta of SN Ic's. These are probably related
to core-collapse events and are known to be hydrogen- and
helium deficient, i.e. their metal content is enhanced. Strong 
iron emission features in the early optical spectra, near the 
peak of the light curve, of this type of supernovae 
might indicate that in particular the iron abundance
is significantly higher than in ordinary stellar atmospheres or
in the ISM. Since very little is known about the actual relative 
element abundances in the ejecta, and the current X-ray data of
GRB afterglows do not allow a composition analysis from the 
observational point of view, I will assume that the torus
material has a tenfold iron overabundance with respect to 
the abundances quoted above, unless explicitly stated otherwise.

\section{Model setup for GRB~970508}

GRB~970508 was a moderately bright burst with a peak flux of 
$\Phi_{\rm p} \approx 3.4 \cdot 10^{-7}$~ergs~cm~s$^{-2}$ and a 
duration of $t_{\gamma} \approx 15$~s in the energy band of the GRBM 
on board the {\it BeppoSAX} satellite (40 -- 700~keV). The X-ray flux 
measured by the WFC of {\it BeppoSAX} exhibits a power-law decay 
($F_{\nu} (t) \propto t^{-\chi}$) with index $\chi = 1.17 \pm 0.1$, 
until the time of a secondary X-ray outburst at $t \sim 6 \cdot 
10^4$~s after the GRB (\cite{piro98}). Piro et al. 
(\markcite{piro99}1999) have recently reported the marginal 
detection of a possible Fe~K$\alpha$ line with a line flux of 
$\Phi_{\rm L} = (5 \pm 2) \cdot 10^{-5}$~cm$^{-2}$~s$^{-1}$ at 
the likely redshift of the burst, $z = 0.835$ (\cite{metzger97}), 
during the first segment of the first period of observation 
by the NFI of {\it BeppoSAX} ($2.0 \cdot 10^4$~s -- $5.6 \cdot 
10^4$~s from the burst trigger time). The end point of 
this time slot was chosen as the onset of the secondary 
X-ray outburst. 

When estimating the efficiency of reprocessing the illuminating
GRB flux into Fe~K$\alpha$ line flux and thus estimating the
amount of mass required to produce the observed fluorescence
line, it is important to take the effects of anisotropy of
the CBM into account, as described in the previous section.
If the density anisotropy due to the torus is strong, the
assumption of the dense torus being illuminated and photoionized
by radiation with the observed GRB characteristics will obviously 
yield a completely unrealistic picture, unless the blast wave is
significantly decelerated by the more dilute material inside the 
radius of the torus before interacting with the torus. The latter
condition would be equivalent to the torus being located outside
the deceleration radius of the blast wave in the dilute external
medium, i.e. $r_{\rm T} \gg r_{\rm d}^{\rm ISM} = [3 \, E_0 / 
(4 \pi \, n_{\rm ISM} \Gamma_0^2 \, m_{\rm p} c^2)]^{1/3} = 1.2 
\times 10^{17} \, [E_{54} / (n_0 \Gamma_{300}^2)]^{1/3}$~cm, 
where $n_0 = n_{\rm ISM} / (1 {\rm cm}^{-3})$. If this 
is not the case, the material in the dense torus is expected to 
be illuminated by radiation, the photon spectrum and temporal 
evolution of which is drastically different from the observed 
GRB spectrum. This is obvious from Eqs. \ref{xd} to \ref{dt_t}. 
Although in this paper the external shock model is used to
parametrize the GRB radiation and blast wave evolution, the
same conclusions are true for an internal shock model for the
prompt GRB radiation.

The duration of the observed line emission is most probably 
dominated by light-travel time effects rather than the intrinsic 
time scale of one of the physical process involved, such as 
recombination or electron-collisional ionization in a hot plasma, 
if the GRB and its afterglow are related to a relativistically 
expanding blast wave. This is because any point at a distance 
$R = 10^{16} \, R_{16}$~cm from the center of the burst souce 
will either be swept up by the blast wave soon after the first 
light front of the GRB has reached it, or the blast wave will be 
effectively stopped before it reaches this point. During the 
coasting phase of the blast wave evolution, it is only 
illuminated for a time period

\begin{equation}
t_{ill} = {R \over \beta_{\Gamma} \, c} \, (1 - \beta_{\Gamma})
\approx 1.9 \> {R_{16} \over \Gamma_{300}^2} \> {\rm s},
\label{till}
\end{equation}
where $\beta_{\Gamma} = \sqrt{1 - \Gamma^{-2}}$. On the other hand,
if this segment of the GRB environment is misaligned with respect
to our line of sight to the GRB, then any isotropically emitted
radiation from there will reach the observer with a time delay
$\Delta t \sim R / c \sim 3 \cdot 10^5 \, R_{16} \, {\rm s}
\gg t_{ill}$. 

A pessimistic estimate of the necessary mass is based on the
result of B\"ottcher et al. (\markcite{boettcher99a}1999a) that out 
to a radius of $R \sim 10^{20} \, (n_{\rm T} [{\rm cm}^{-3}])^{-1/3}$~cm 
iron in the CBM will be completely ionized. In a scenario in which
the density of the illuminated material is relatively low so that
recombination is negligible, the Fe~K$\alpha$ line can only be
emitted via fluorescence. Taking into account the large number 
of Auger electrons ejected following photoionization of iron in 
low ionization states, a fiducial number of K$\alpha$ line photons 
in the energy range 6.4 -- 6.7~keV emitted in the course of complete 
ionization of an iron atom is $\sim 5$ since it takes on average
$\sim 12$ X-ray photons to ionize an initially neutral Fe atom
completely (\cite{weisheit74}), and the K$\alpha$ fluorescence
yields for the various Fe ions are typically in the range 0.3 -- 0.4. 
Thus, for a line luminosity $L_{K\alpha} = 10^{44} L_{44}$ergs~s$^{-1}$ 
emitted over a time scale $\Delta t_{\rm L} = 10^5 \, t_5$~s, a total of

\begin{equation}
N_{\rm Fe} = 2 \cdot 10^{56} \, {L_{44} \, t_5 \over f}
\label{NFe}
\end{equation}
iron atoms is needed, where $f \ge 1$ is a correction factor 
accounting for the enhancement of the efficiency of line emission
due to recombination and electron-collisional ionization. Eq.
\ref{NFe} yields a required mass of iron of

\begin{equation}
M_{\rm Fe} = 0.16 \, {L_{44} \, t_5 \over f} \> M_{\odot}.
\label{MFe}
\end{equation}
The line flux measured in the afterglow of GRB~970508 leads 
to an estimated mass of $M_{Fe} \sim 1 \> M_{\odot}$ of iron,
concentrated in a region of $\lesssim 10^{-3}$~pc, which is 
highly unlikely to be realized in any known and well-understood 
astrophysical environment.

Assuming a dominant role of the effects of recombination and electron-impact
ionization, one finds a more realistic lower limit on the required iron
mass of $M_{\rm Fe} \gtrsim 6 \cdot 10^{-6} \> M_{\odot}$ (\cite{lcg99}).
It may thus be concluded that the effects of recombination and
collisional ionization are dominant in the formation of the
Fe~K$\alpha$ line in GRB afterglows. In order for recombination
to be efficient, the recombination time scale of iron needs to
be $t_{rec} \ll 10^5$~s. Assuming temperatures of $T \sim
10^7$ -- $10^8$~K of the torus material, the typical recombination
coefficients of highly ionized iron ions are $\alpha_{rec} \sim
10^{-11}$~cm$^3$~s$^{-1}$, which implies that the density of
material in the torus must be $n_T \gg 10^6$~cm$^{-3}$.

In the framework of the analytical representation of Dermer et al.
(\markcite{dcb}1999), the spectral and temporal properties of 
GRB~970508 can be reasonably well reproduced assuming a total
energy of the explosion of $\partial E / \partial\Omega 
= 3.6 \cdot 10^{52}$~ergs/sr, an initial bulk Lorentz factor 
of $\Gamma_0 = 100$, a uniform density of the ISM of $n_{\rm ISM}
= 4.8 \cdot 10^5$~cm$^{-3}$, an equipartition factor $q = 8.15 
\cdot 10^{-5}$, a quasi-adiabatic blast wave with $g = 1.6$ 
(the blast wave slows down as $\Gamma(r) \propto r^{-g}$ 
outside the deceleration radius), a low-energy spectral 
index (photon number index) of $\alpha_{\rm le} = 2/3$ and a 
high-energy index of $\alpha_{\rm he} = 2.1$. These parameters
yield a deceleration time scale (which may be identified with the
burst duration at $\gamma$-ray energies) of $t_{\rm d} = 15$~s and
a temporal decay of the X-ray afterglow with index $\chi = 1.2$.
The peak of this model $\nu F_{\nu}$ spectrum during the prompt GRB 
phase is at $E_{\rm pk} = 45$~keV, and its isotropic $\nu L_{\nu}$ peak
luminosity at this photon energy is $2.8 \cdot 10^{50}$~ergs~s$^{-1}$.
All these features of the model GRB spectrum and spectral evolution 
are in good agreement with the observed properties of GRB~970508
(\cite{piro98}). Obviously, the secondary X-ray flare at $t
\approx 6 \cdot 10^4$~s is not modelled with this approach.

According to the synchrotron-shock model, the temporal decay 
index $\chi$ of the X-ray afterglow (produced by strongly 
cooled electrons behind the shock) is related to the 
radiative-regime parameter $g$, the decay index $\eta$ of 
the external matter density ($n_{\rm ISM} \propto r^{-\eta}$), 
and the high-frequency $\nu F_{\nu}$ spectral index $\delta = 
\alpha_{\rm he} - 2$ through

\begin{equation}
\chi = {4 g (1 + \delta) + \eta (1 + \delta/2) - 2 \over
2 g + 1}
\label{chi}
\end{equation}
(\cite{dcb99}). Solving this for $\eta$, inserting $\chi = 1.17$, 
keeping in mind that $g \ge 1.5$, and assuming $\delta > 0$, we 
find that the external matter density gradient can not be steeper 
than $\eta \le 0.68$. This means that within the framework of 
the standard synchrotron-shock model, the observed afterglow can
not have been produced in an isotropic stellar-wind environment 
of the putative stellar progenitor, for which one would generally
expect an $r^{-2}$ density profile.

In the simulations presented here, the GRB explosion is allowed
to be intrinsically anisotropic. In order not to introduce an 
uncontrollably large parameter space, only the effects of an
angle-dependent energy output $(\partial E / \partial \Omega)
(\theta)$ and baryon loading (or, equivalently, bulk Lorentz 
factor) $\Gamma (\theta)$ are investigated in this paper. 

\section{Results}

In the first step, a series of simulations with intrinsically
isotropic blast waves as specified in the previous section has 
been carried out in order to specify appropriate torus parameters 
to produce an iron line luminosity of $\sim 10^{44}$~ergs/s. In
all simulations shown here, the torus is assumed to have an average 
distance of $r_{\rm T} = 7 \cdot 10^{-4}$~pc from the center 
of the explosion in order to produce the observed time delay
of the iron line emission with respect to the prompt GRB phase. 
As motivated in \S 2, the material in the torus has a tenfold 
overabundance of iron with respect to standard solar-system 
abundances, which are assumed for the surrounding ISM. 

\begin{figure}
\epsfysize=11cm
\rotate[r]{
\epsffile[50 70 550 500]{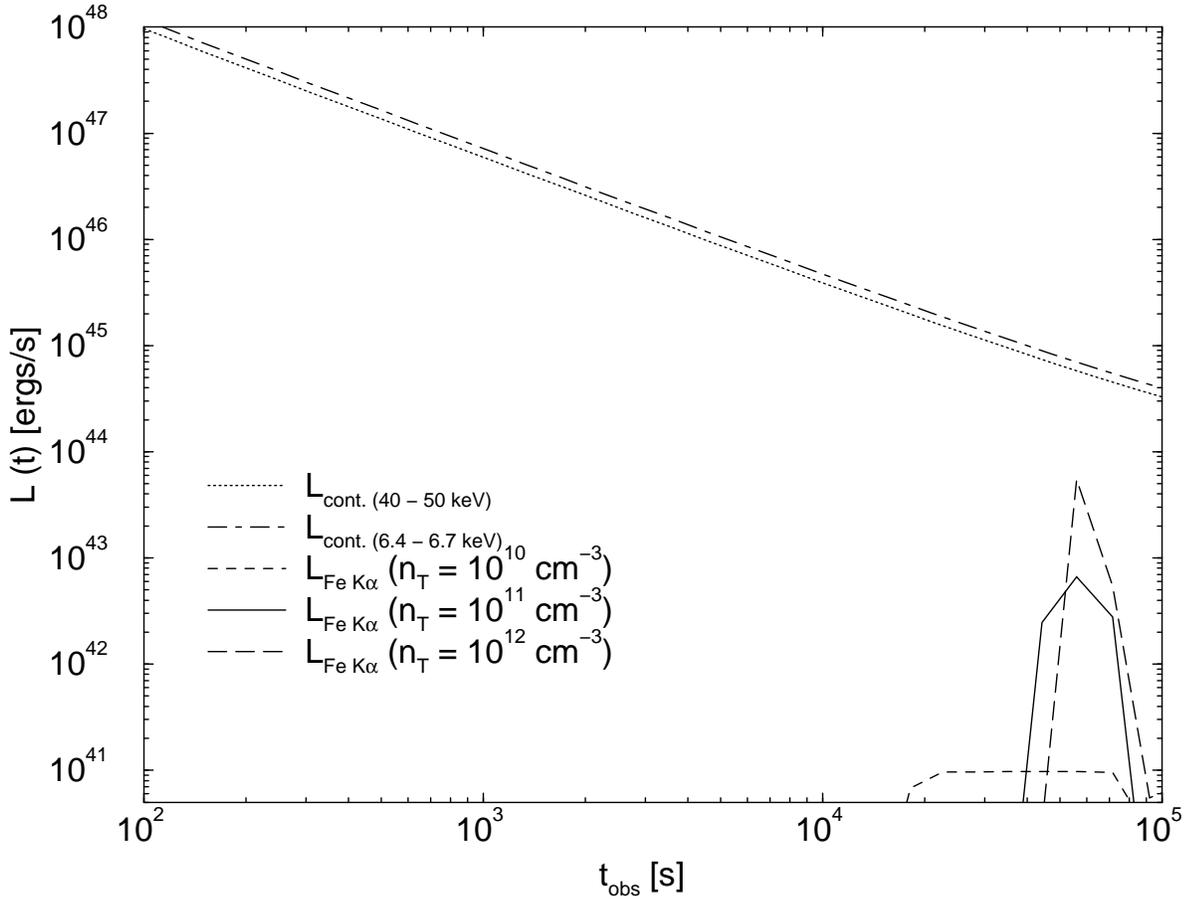}}
\caption[]{Light curves of the apparent isotropic luminosity in two
continuum energy bands (dotted and dot-dashed curves, respectively)
and in the Fe~K$\alpha$ line for three different values of the density
$n_{\rm T}$ of torus material. All other parameters are fixed (see text).
The blast wave is assumed to be intrinsically isotropic; the torus
material has a tenfold overabundance of iron.}
\label{density}
\end{figure}

Fig. \ref{density} illustrates the dependence of the produced 
iron line on the density of torus material, while the total 
mass of material in the torus is held constant (i.e. $a 
\propto n_{\rm T}^{-1/2}$). The iron line intensity is strongly 
positively correlated with the density, which indicates the 
dominance of the density-dependent processes like electron-impact 
ionization and recombination in the formation of the line. 
Given a total torus mass of $M_{\rm T} \lesssim 1 \> M_{\odot}$ 
and the assumed tenfold iron overabundance, a density of 
$n_{\rm T} \gtrsim 10^{12}$~cm$^{-3}$ is necessary to produce 
the observed Fe~K$\alpha$ line intensity. The figure also 
shows that the duration of the flash of iron line emission 
is determined by light travel time effects rather than the
time scale of elementary processes because due to the high
densities in the torus the recombination and collisional
ionization time scales are $\ll 10^4$~s and thus shorter
than the light travel time through the torus itself. 

\begin{figure}
\epsfysize=11cm
\rotate[r]{
\epsffile[50 70 550 500]{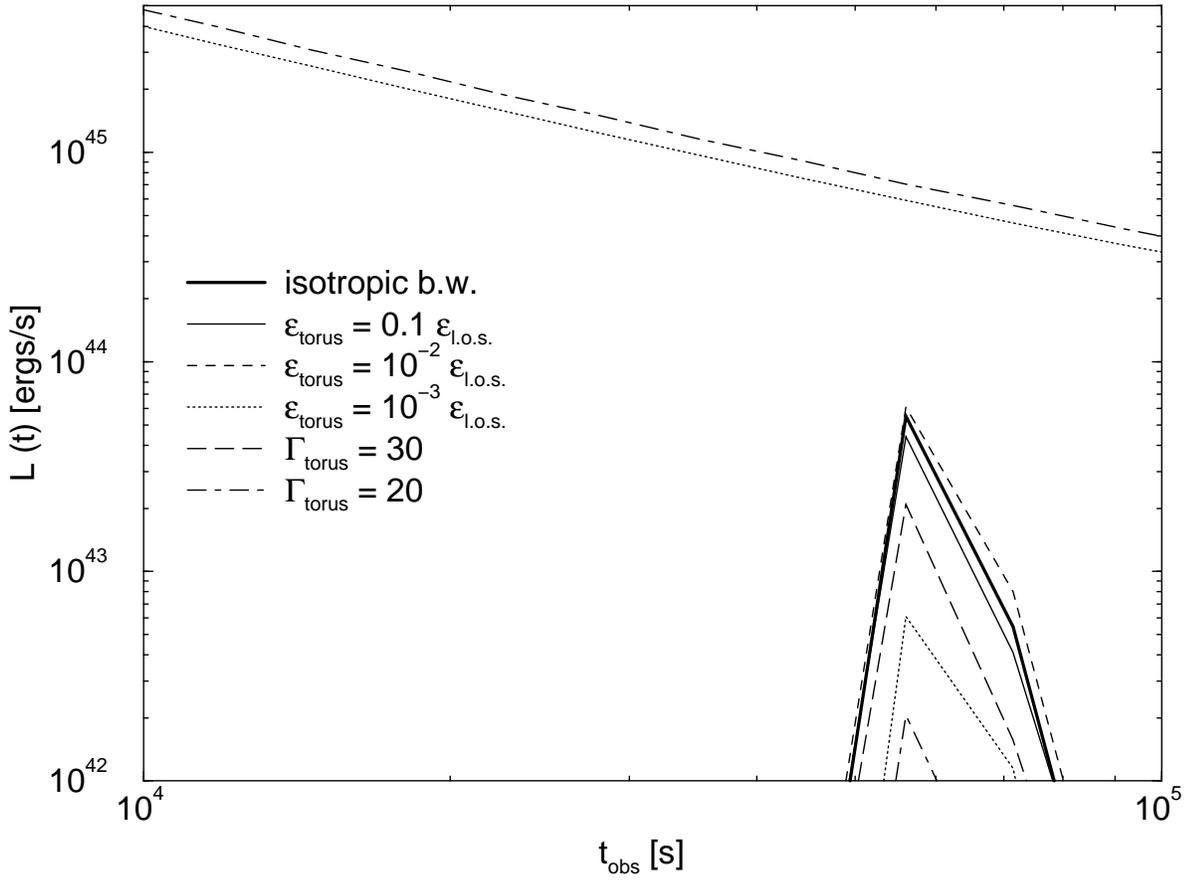}}
\caption[]{The light curves of the Fe~K$\alpha$ line emission for
an observer located along the symmetry axis ($\theta = 0$) compared
to the on-axis light curves of the continuum, for an intrinsically 
anisotropic blast wave. In the legend, $\epsilon \equiv \partial
E / \partial\Omega$. }
\label{anisotropy}
\end{figure}

In Fig. \ref{anisotropy}, the effects of an intrinsically
anisotropic blast wave are illustrated. In accord with the
previous result that it is mainly electron-impact ionization
and recombination which is responsible for the iron line
emission, the effect of a limited amount of intrinsic anisotropy 
of the blast wave are only minor, as long as $[\partial E / \partial
\Omega ({\rm torus})] \gtrsim 0.01 [\partial E / \partial
\Omega ({\rm l.o.s.})]$ and $\Gamma \gtrsim 30$. This is because 
the torus material rapidly becomes highly ionized and cools 
down to temperatures favorable for the production of an 
Fe~K$\alpha$ line on time scales of $\ll 10^4$~s, only
weakly dependent on the actual amount of energy deposited
and on the detailed spectral characteristics of the illuminating
radiation. The simulations show that the material ahead of the
non-relativistic shock wave propagating through the torus is
efficiently heated by the intense X-ray and gamma-ray flash
produced during the initial deceleration of the blast wave 
within the torus. In these simulations, $[\partial E / \partial
\Omega ({\rm torus})] < [\partial E / \partial \Omega ({\rm l.o.s.})]$ 
with $\Gamma_{\rm torus} = \Gamma_{\rm l.o.s.}$ corresponds to
a smaller amount of mass ejected in the direction of the torus
than along the evacuated funnel which points towards the observer.
The case $\Gamma_{\rm torus} < \Gamma_{\rm l.o.s.}$ with $[\partial 
E / \partial \Omega ({\rm torus})] = [\partial E / \partial 
\Omega ({\rm l.o.s.})]$ corresponds to a higher baryon contamination
in the direction of the torus.

\begin{figure}
\epsfysize=11cm
\rotate[r]{
\epsffile[50 70 550 500]{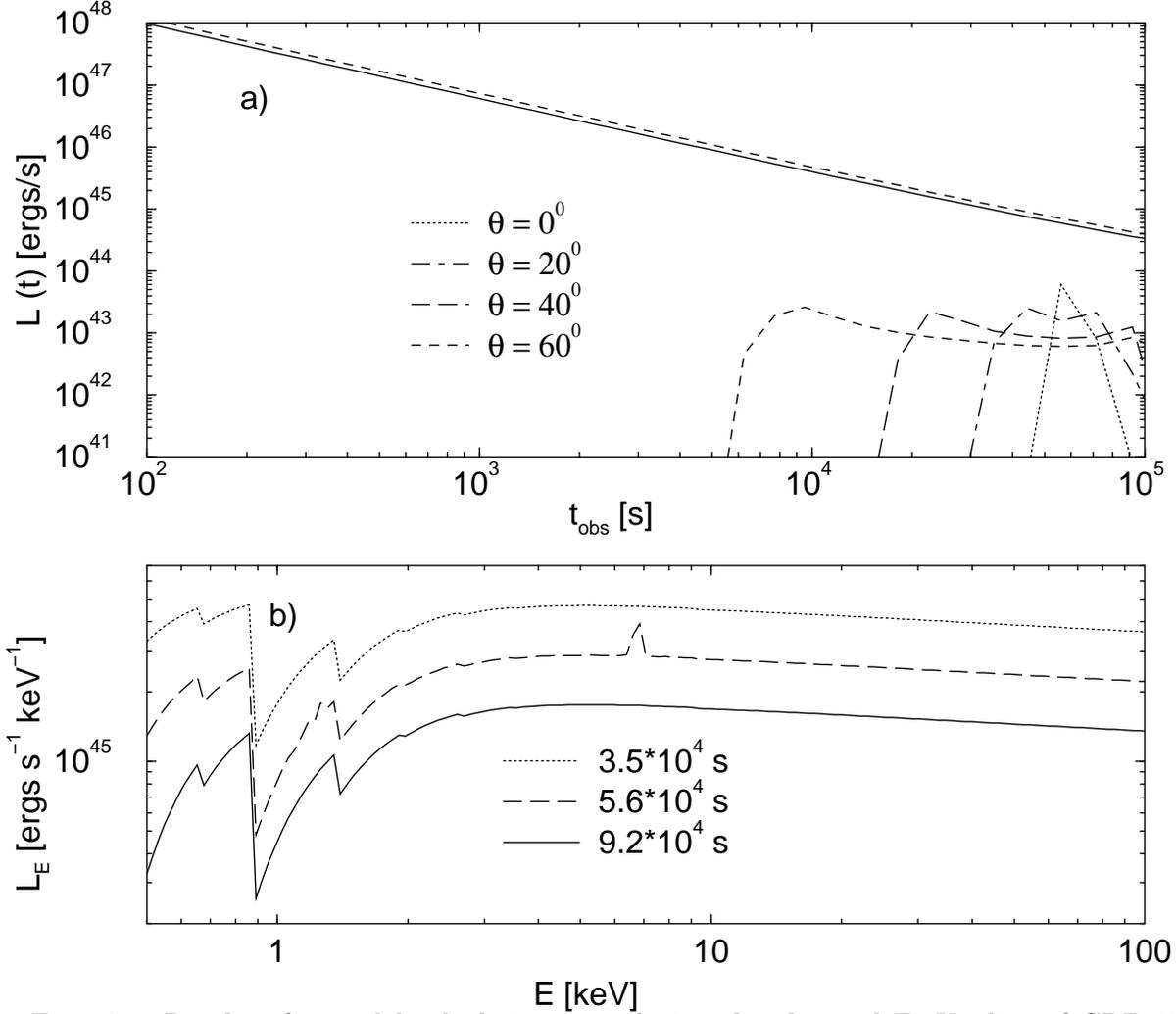}}
\caption[]{Results of a model calculation reproducing the
observed Fe~K$\alpha$ line of GRB~970508. The torus has a
density of $n_{\rm T} = 10^{12}$~cm$^{-3}$ and a total mass of
$M_{\rm T} = 0.71 \> M_{\odot}$. The torus material is iron-enriched
by a factor of 10 w.r.t. standard solar-system abundances,
corresponding to a total mass of iron of $M_{\rm Fe} = 2.2 \cdot 
10^{-4} \> M_{\odot}$ in the torus. The blast wave is
anisotropic with $[\partial E / \partial\Omega ({\rm torus})]
= 10^{-2} \cdot [\partial E / \partial\Omega \, ({\rm l.o.s.})]$.
The upper panel (a) shows the light curves in the iron 
line emission for different observing angles compared 
to the decay of the continuum at 40 -- 50~keV (solid) 
and at 6.4 -- 6.7~keV (short-dashed). 
The lower panel (b) shows the simulated energy spectra for 
an observer located along the symmetry axis ($\theta = 0^o$) 
at different times after the GRB.}
\label{best_fit}
\end{figure}

Fig. \ref{best_fit} shows the result of a simulation which 
yields a reasonable representation of the afterglow spectrum
of GRB~970508. Panel \ref{best_fit}a shows the angle-dependent
apparent luminosity of the Fe~K$\alpha$ line compared to the
luminosity in two continuum energy bands as a function of time.
Panel \ref{best_fit}b shows the simulated energy spectra at 3 
different times for an observer located on the symmetry axis.
The simulation predicts that the Fe~K$\alpha$ line emission is
only emitted in a rather short flash of a few times $10^4$~s 
if the burst is observed close to the symmetry axis of 
the system. Note that the absorption features visible in
Fig. \ref{best_fit} are due to the influence of the ISM rather
than the torus. However, as has been shown by B\"ottcher et al. 
(\markcite{boettcher99a}1999a) fluorescence line emission associated
with this photoelectric absorption within the quasi-isotropic ISM
may emerge above the background of the afterglow continuum only 
several weeks or months after the burst and is expected to have 
too low a flux level to be observable by any existing or currently 
planned X-ray telescope.

\begin{figure}
\epsfysize=11cm
\rotate[r]{
\epsffile[50 70 550 500]{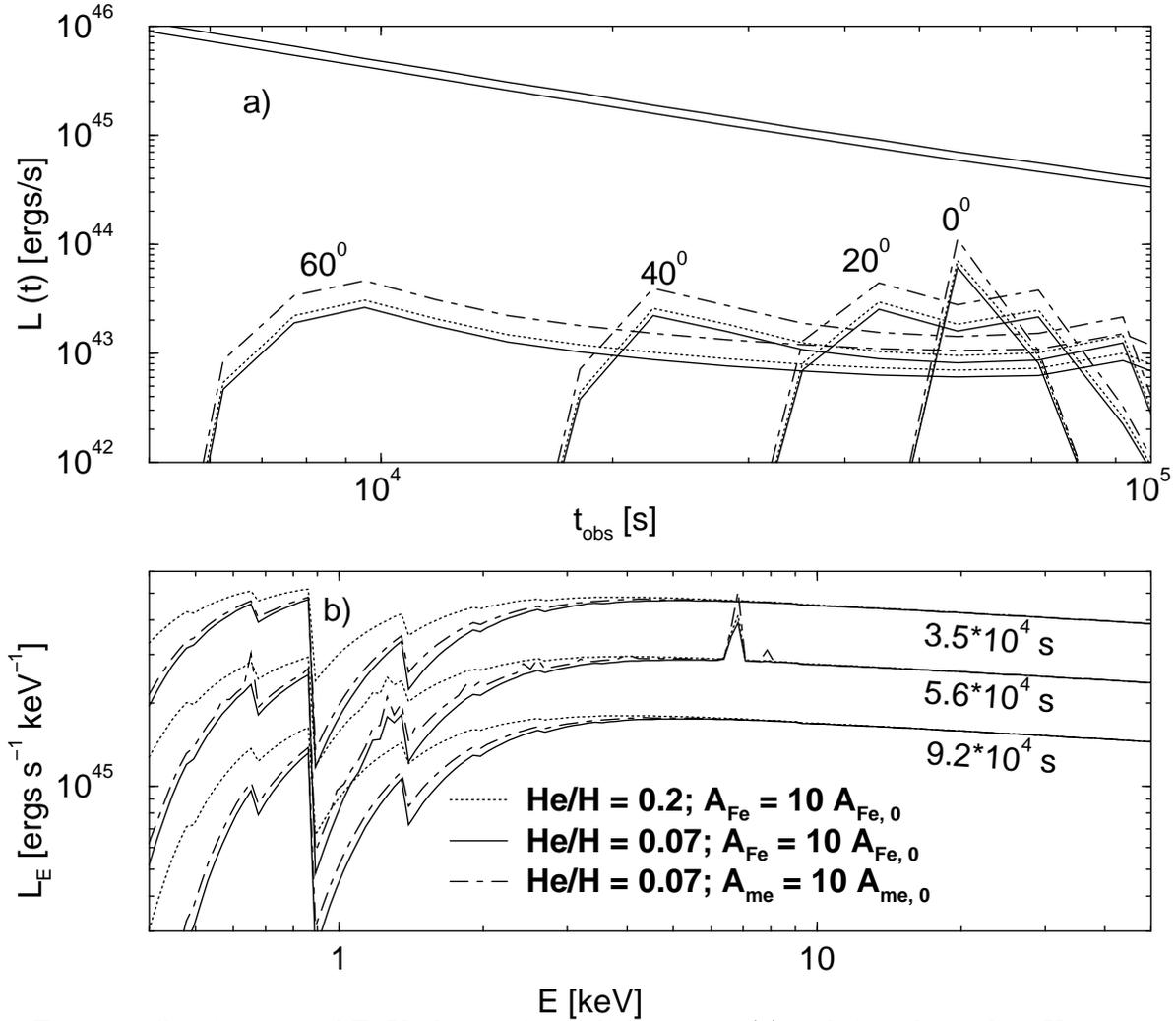}}
\caption[]{Continuum and Fe~K$\alpha$ line emission light curves (a) and 
time-dependent X-ray spectra (b) for the case shown in Fig. \ref{best_fit}
(solid curves), compared to test cases with (dotted curves:) an enhanced 
He/H ratio of 0.2 (default value: 0.07), and (dot-dashed curves:) with 
tenfold overabundance of all elements heavier than helium (default case: 
only iron enhanced). }
\label{abundances}
\end{figure}

The actual composition of the material in the torus is rather
uncertain. Fig. \ref{abundances} illustrates how the predicted light
curves and time-dependent X-ray spectra change for different He/H
ratios and metal enhancement properties. The dotted curves show a 
test case with a He/H ratio of 0.2 (compared to the default value 
of 0.07), while the dot-dashed curves illustrate a case in which the 
abundances of all metals (i.e. all elements heavier than helium) is 
enhanced by a factor of 10. The expected equivalent width of the
Fe~K$\alpha$ line depends only weakly on the He/H ratio through the
slightly higher electron density (and thus higher electron-impact
ionization rate) in the helium-enhanced, highly ionized torus 
with respect to the standard He/H ratio. However, this also leads
to more intense bremsstrahlung and and radiative-recombination continuum
emission from the hot, highly ionized torus, which shows up at energies
$\lesssim$~a few keV. Since this radiation propagates toward the observer
through already highly ionized portions of the ISM, this continuum
emission from the torus leads to an apparent weakening of the absorption
features imprinted onto the afterglow radiation by the ISM. 

An overabundance of all metals in the torus material (see dot-dashed
curves in Fig. \ref{abundances}), leads to generally stronger emission
line features around $t \sim 6 \times 10^4$~s. In addition to the 
Fe~K$\alpha$ line, now also a weak Ni~K$\alpha$ line at 7.5 -- 7.7~keV 
and weak K$\alpha$ fluorescence and recombination lines from sulfur
(2.3 -- 2.4~keV), silicon (1.7 -- 1.8~keV), and magnesium ($\sim 1.3$~keV),
and an Fe~L fluorescence line at $\sim 0.7$~keV become visible in the
spectrum for a period of a few $\times 10^4$~s.

\section{Summary and conclusions}

I have presented results of a detailed numerical study of a 
scenario in which a relativistic blast wave associatd with a
cosmological gamma-ray burst irradiates and interacts with a
torus of dense, metal-enriched material surrounding the burst 
source. The effects of photoionization, fluorescence line 
emission, recombination, electron-impact ionization, 
bremsstrahlung, Compton scattering, Coulomb scattering,
and the hydrodynamical interaction of the blast wave with the
torus have been considered. The effects of variations of
several model parameters have been investigated, and a model 
calculation to reproduce the marginal detection of an 
Fe~K$\alpha$ line in the X-ray afterglow of GRB~970508 
has been presented.

The results indicate that in order to reproduce the observed
iron line in the afterglow of GRB~970508 (if real), an iron mass
of $\sim 10^{-4} \, M_{\odot}$ is required to be concentrated in
a region of dense ($n_{\rm T} \gtrsim 10^{12}$~cm$^{-3}$) material. 
The line is predominantly produced by electron-impact ionization 
and recombination in the hot ($T \sim 10^8$~K), highly ionized 
torus. In order to give a realistic description of the model 
scenario investigated here, it is important to consider the effect 
of the anisotropy of the GRB environment on the radiation effectively
illuminating the dense material. Unless the blast wave is substantially
decelerated by the dilute ISM before interacting with the torus, the
spectrum and spectral evolution of the illuminating radiation will be
drastically different for the torus material with respect to the 
observed characteristics of the GRB and afterglow radiation.

The amount of line emission is rather weakly dependent on the
He/H ratio in the torus material. However, significant deviations
in the abundances of heavier elements from the standard solar-system
values could be detectable through the respective fluorescence and
recombination lines from these elements. 

Given the limited significance of the detections of the iron lines 
in both the afterglows of GRB~970508 an GRB~970828, rapid follow-up
observations by more sensitive X-ray telescopes with high spectral
resolution are necessary to confirm the reality of these spectral
features. Modern X-ray satellites like {\it Chandra} and {\it XMM} 
will be able to perform these observations with unprecedented 
sensitivity and spectral resolution. In case of positive detections,
these X-ray line emission and absorption features may be an extremely 
powerful tool for redshift measurements and probing the environments 
of cosmological GRBs. This could ultimately help to distinguish between 
the two basic classes of GRB progenitor models, i.e. models related to 
the deaths of supermassive stars versus models involving the mergers 
of compact objects.

An interesting conclusion from these simulations is that X-ray 
flashes due to bremsstrahlung and radiative recombination continuum 
radiation from the dense, shock-heated torus should occasionally be 
observable even if the GRB itself is misaligned with respect to 
the observing direction. Such X-ray flashes with strong Fe~K$\alpha$ 
line features may be hard to detect in archival data from, e. g.,
{\it ASCA} or {\it BeppoSAX}, but the increased sensitivity and 
imaging capability of {\it XMM} and {\it Chandra} may enable their 
detection.

\acknowledgements{The work of M.B. is supported by NASA through
Chandra Postdoctoral Fellowship Award Number PF~9-10007, issued 
by the Chandra X-ray Center, which is operated by the Smithsonian 
Astrophysical Observatory for and on behalf of NASA under contract 
NAS~8-39073. I thank J. M. Shull for communicating the corrected 
values of the recombination rate coefficients to me, and E. P. Liang 
for stimulating discussions and careful reading of the manuscript.
I also thank the anonymous referee for a very detailed review and
many useful comments.}

\newpage

\newpage

\newpage

\newpage

\newpage

\end{document}